\documentstyle[prd,aps,psfig]{revtex}
\begin{document}
\draft
\title{Cosmological gravitons and the expansion dynamics during the matter age}
\author{M. R. de Garcia Maia\thanks{Electronic address: mrgm@dfte.ufrn.br},
J. C. Carvalho\thanks{Electronic address: carvalho@dfte.ufrn.br}, and
J. S. Alcaniz\thanks{Electronic address: alcaniz@dfte.ufrn.br}}
\address{Departamento de F\'{\i}sica, Universidade Federal do Rio
Grande do Norte, 59072-970 Natal RN Brazil}
\maketitle
\begin{abstract}
Extending previous results [Phys. Rev.D {\bf 56}, 6351 (1997)], we
estimate the back reaction of cosmological gravitons in the expansion dynamics,
during the matter age. Tensor perturbations with scales larger than the Hubble
radius are created due to the inequivalence of vacua at different times.
During noninflationary phases these perturbations become effective
gravitational waves as they enter the Hubble radius,
adding new contributions to the energy density of the subhorizon
waves. During the radiation epoch the creation of these effective
gravitons may lead to a departure from the standard behaviour
$a(t)\propto t^{1/2}$, with possible consequences to several cosmic processes, such as
primordial nucleosynthesis. We examine the implications of this
phenomenom during the matter dominated era, assuming an initial inflationary
period. The dynamical equation obeyed by the scale factor is
derived and numerically solved for different values of the relevant
parameters involved.
\end{abstract}
\section{INTRODUCTION}
\label{s1}
In a recent paper \cite{gw6}, we have studied the back reaction of cosmological gravitons
on the  cosmic dynamics during the radiation epoch. The analysis was based on the
following reasoning: the inequivalence of vacuum states at different moments of time
leads to the production of tensor perturbations in scales larger than the Hubble length
[1 -- 8].
During noninflationary periods of expansion, these very
long tensor perturbations (VLTP's) become effective gravitational waves (EGW's) as they
enter the Hubble radius, thus adding new contributions to the energy density associated
with the subhorizon waves $\rho_g$ [1 -- 5], \cite{zeldovich}.
Such a
phenomenum can be studied as a process of creation of effective gravitons by using the
macroscopic formalism to matter creation based on the thermodynamics of open systems
\cite{prigogine,calvao}. In order to deal with the process, a creation pressure term
is introduced in the continuity equation obeyed by $\rho_g$. A dynamical equation for
the scale factor $a(t)$ is then derived which takes into account the effective
gravitons'
back reaction. Full details of this approach can be found in  Ref. \cite{gw6} where the
equation for $a(t)$ was numerically solved for a model in which the universe evolves
from an arbitrary initial phase to a radiation dominated era. It was found that, if the
barotropic index $\gamma$ of the equation of state in the first epoch is close to $2/3$,
the back reaction of the effective gravitational waves makes $a(t)$ deviate appreciably from
the standard behavior $a(t) \propto t^{1/2}$.

In the present paper we extend our analysis to the matter dominated period of the cosmic
expansion. In particular, we aim to check a conjecture stated by Sahni
\cite{sahni},
according to which  the process mentioned above would
ultimately lead the universe to expand linearly with time ($a(t) \propto t$), since the
tensor perturbations would enter the Hubble radius during noninflationary periods of
expansion (increasing $\rho_g$) and leave the Hubble radius during inflationary periods
(decreasing $\rho_g$). In Ref. \cite{gw6} we have shown that the scale factor would turn
from $a(t) \propto t^{1/2}$ to $a(t) \propto t$ only if $\gamma \rightarrow 2/3$.
Several models with the universe evolving linearly with time during its early
phases have been proposed, most of them related to string
motivated cosmologies \cite{vilenkin84,gasperini92}.
An universe evolving linearly with time during most of its history would lead to important
observational consequences, such as those related to the age of the
universe, to the luminosity distance - redshift and the angular
diameter distance - redshift relations, and to the galaxy number
count as a function of the redshift \cite{kolb89,akdeniz94} (see also
\cite{wesson92}).

We should remark that, in order to the mechanism of graviton creation take place, no
`exotic' physics has to be assumed. It only requires the validity of quantum mechanics and
of general relativity \cite{gri93}. Therefore, the study of the possible influence of the
cosmological gravitational waves in the expansion dynamics has an importance on its own,
a fact that has not been fully appreciated in the literature.

The paper is organized as follows: In Section \ref{s2}
we present the basic equations related to
the spectrum of cosmological gravitational waves. In Section \ref{s3}
we derive the dynamical
equations that govern the expansion dynamics during the matter age, taking into account the
gravitons' back reaction. In Section \ref{s4} we show the numerical solutions of these
equations and present our conclusions.

The system of units used is such that $\hbar = c = k_B =1$.
\section{BASIC EQUATIONS}
\label{s2}
We will consider a homogeneous and isotropic universe, for which the background line element
takes the Friedmann-Robertson-Walker (FRW) form
\begin{equation}
\label{e1}
ds^2 = dt^2 - a^2(t)dl^2 = a^2(\eta)(d\eta^2 -dl^2)\;,
\end{equation}
where $t$ and $\eta$ are, respectively, the cosmic and conformal times, related
by
\begin{equation}
\label{e2}
dt=ad\eta\;.
\end{equation}
We will restrict ourselves to the spatially flat case. If the pressure $p$ and energy density
$\rho$  of the cosmic fluid are related by the equation of state
\begin{equation}
\label{e3}
p=(\gamma -1)\rho\;,
\end{equation}
then the scale factor is \cite{gw1}
\begin{equation}
\label{e4}
a(t)= a_0\left [ 1+\frac{3\gamma H_0}{2}(t-t_0)\right ]^{2/(3\gamma)}\;,
\end{equation}
where $H_0 \equiv H(t_0)$ and $H\equiv \dot{a}/a$.

We will further assume that the universe evolves from an initial arbitrary era ($\gamma =
\gamma$) to a radiation dominated phase ($\gamma = 4/3$) and then to a matter dominated period
($\gamma = 1$).

As the inequivalence of vacua appear at different instants of time, we will focus on the
time-dependent amplitude of the tensor perturbations $\mu(k,\eta)$, which obeys
\cite{gw6,gw1,gw3t}
\begin{equation}
\label{e5}
\mu^{''}(k,\eta)+\left (k^2-\frac{a^{''}}{a}\right )\,\mu(k,\eta)=
0\;.
\end{equation}
In the above equation the primes indicate derivatives with respect to the conformal time and
$k$ is the comoving wave number, related to the physical wavelength $\lambda$ and frequency
$\omega$ by
\begin{equation}
\label{e6}
k=\frac{2\pi a}{\lambda} = \omega a\;.
\end{equation}
The properly normalized solutions of Eq. (\ref{e5}), corresponding to a scale
factor given by Eq. (\ref{e4}), and representing adiabatic
vacuum states \cite{birrel,fulling}, can be found in Refs. \cite{gw1,gw3t}.

The spectrum of the EGW's can be described by the quantity
$P_g(\omega)$, defined in such a way that $P_g(\omega)\,d\omega$ represents
the
energy per unit volume between the frequencies $\omega$ and $\omega+d\omega$
\cite{allen,gw1,gw3t}. Note that this quantity can be defined only for the
EGW's \cite{allen,zeldovich}, i.e., for those perturbations such that
\begin{equation}
\label{e7}
\lambda\leq\lambda_H\equiv H^{-1}\;.
\end{equation}
In the units used in this paper and assuming an initial vacuum state, $P_g(\omega)$ is given
by \cite{allen,gw1,gw3t}
\begin{equation}
\label{e8}
P_g(\omega)=\frac{\omega^3}{\pi^2}\,\langle N(\omega)\rangle\;.
\end{equation}
In the above equation $\langle N(\omega)\rangle$ is the
expectation number of the gravitons with frequency $\omega$ which is found to be
\cite{gw6,gw1,gw3t}
\begin{equation}
\label{e9}
\langle N(\omega)\rangle = |\beta(k)|^2\;,
\end{equation}
where $\beta(k)$ and $\alpha(k)$ (see below) are the Bogoliubov coeficients relating the
creation and annihilation operators (which define the particle states) of different epochs.
We will denote by $\alpha_1$, $\beta_1$ the coeficients associated with the first transition
from an arbitrary phase to the radiation dominated era, and by $\alpha_2$, $\beta_2$ those
related with the transition from the radiation to the matter dominated epoch. These
coefficents have been obtained in references \cite{gw1,gw3t} (see also \cite{gw6}).
The coefficient $\beta$ is then given by \cite{allen}
\begin{equation}
\label{e10}
\beta = \beta_2(k)\alpha_1(k)+\alpha_{2}^{*}(k) \beta_1(k)\;,
\end{equation}
and the asterisk indicates the complex conjugate of a quantity.

The total energy density associated with the tensor perturbations is
obtained by integrating $P_g(\omega)$,
\begin{equation}
\label{e11}
\rho_g(t)=\int^{\infty}_{\omega_{min}(t)}P_g(\omega)\,d\omega\;,
\end{equation}
where
\begin{equation}
\label{e12}
\omega_{min}(t)=2\pi H(t)\;.
\end{equation}
This infrared cutoff is obviously related to the condition (\ref{e7}) and,
since it is time dependent, it is the origin of the process of creation of
EGW's. In a FRW scenario, it will be responsible for the fact that $\rho_g$
does not decay with $a^{-4}$, as one could expect for a massless particle such as the
graviton.

The integral in Eq. (\ref{e11}) is usually written with a finite upper limit.
This is done because the above method for obtaining the spectrum, based on
the sudden transition approximation between cosmic eras, does not give accurate results for
high
frequency waves. In fact, it is reliable only for those waves whose
periods are much greater than the transitions time scales \cite{gw3t,gw5}.
However, due to the
adiabatic theorem \cite{fulling}, the number of created
particles should decrease exponentially for the high frequency modes. Hence,
an ultraviolet cutoff is imposed and the error produced by doing so is supposed to be small.
Our present results will not be affected by this approximation since we are obviously
interested only on the very long modes.
\section{DYNAMICAL EQUATIONS DURING THE MATTER ERA}
\label{s3}
Following Ref. \cite{gw6}, we will suppose that, prior to the time $t_1$ the dominant material
content of the Universe
has an equation of state of the form (\ref{e3}).
This stage can be either an inflationary or a noninflationary one.
At $t_1$ a transition occurs, so that the new barotropic parameter is
$4/3$. Due to the transition, tensor perturbations
are created \cite{gw6,gw1,gw3t}. Some
of these have wavelengths less than $H^{-1}(t_1)$ (EGW's), but most are created
with scales larger than $H^{-1}(t_1)$ (VLTP's). A similar phenomenom will occurr
at a time $t_2$ when the universe becomes matter dominated ($\gamma =1$). We will
further assume that during this matter dominated period and until
a time, say, $t_{g}$, the energy density $\rho_g$ associated with the
EGW's
is negligible compared with the energy density of matter $\rho_m$. There is no
restriction over the size of
the interval $t_{g}-t_2$, that can be taken to be arbitrarily small.
However, after $t_{g}$, the ongoing transformation of
VLTP's into EGW's makes the continuous increase of $\rho_g$ start perturbing the dynamics.
Thereafter, the Universe can be supposed to be
filled by two fluids: dust and the one composed by
the effective gravitons. The {\em total} energy
density is then $\rho_m+\rho_g$.
As gravitons require extremely high energies to interact with matter
\cite{gw3t,kT}, the two fluids can be safely considered to be noninteracting.
The effective gravitons behave as a perfect fluid with equation
of state \cite{carr,gri}
\begin{equation}
\label{e13}
p_g=\frac{\rho_g}{3}\;.
\end{equation}
Nevertheless, it is possible to associate a creation pressure term $\Pi_g$ to
the creation process of these gravitons \cite{gw6}. The field equations are then written as
\begin{equation}
\label{e14}
8\pi G\,(\rho_m+\rho_g)=3\frac{\dot{a}^2}{a^2}\;,
\end{equation}
\begin{equation}
\label{e15}
8\pi G\,(p_m+p_g+\Pi_g)=-2\frac{\ddot{a}}{a}-\frac{\dot{a}^2}{a^2}\;,
\end{equation}
\begin{equation}
\label{e16}
\dot{n_g}+3Hn_g=\Psi_g\;,
\end{equation}
where the dots indicate derivatives with respect to the cosmic time, $p_m =0$ is the
pressure of the dust fluid and $n_g$ and $\Psi_g$ are the number density and creation
rate of the effective
gravitons, respectively. This last equation is the novel aspect introduced by the
phenomenological formalism to particle (in this case, gravitons) creation
\cite{prigogine,calvao}. The conservation of the energy-momentum tensor leads to
\begin{equation}
\label{e17}
\dot{\rho}+3H(\rho + p + \Pi_g)=0\;,
\end{equation}
where $\rho$ is the total energy density ($\rho= \rho_m +\rho_g$) and $p$ is the
total pressure ($p=p_m +p_g$, $p_m=0$).

As the two fluids are noninteracting, this conservation equation can be split
into
\begin{equation}
\label{e18}
\dot{\rho_m}+3H\rho_m=0
\end{equation}
and
\begin{equation}
\label{e19}
\dot{\rho}_g+3H\left (\frac{4}{3}\rho_g+\Pi_g\right)=0\;.
\end{equation}

From Eqs. (\ref{e14}), (\ref{e15}), (\ref{e18}) and (\ref{e19}) we obtain \cite{gw6}
\begin{equation}
\label{e20}
\frac{\ddot{a}}{a}+\frac{1}{2}\,\frac{\dot{a}^2}{a^2}=
4\pi G\,\left(\rho_g+\frac{\dot{\rho}_g}{3H}\right)\;,
\end{equation}
\begin{equation}
\label{e21}
\frac{\dot{a}^2}{a^2}=\frac{8\pi G}{3}(\rho_m + \rho_g)\;.
\end{equation}
It is important to mention that, in the method described in \cite{gw6,gw1,gw3t}, the
Bogoliubov coefficients that appear in the calculus of $\langle N(\omega)\rangle$ are
evaluated at the transition times $t_1$ and $t_2$. Hence, $\rho_g(t)$ is univocally
determined in terms of $a(t)$ and $\dot{a}(t)$, that is \cite{gw6}
\begin{equation}
\label{e22}
\rho_g(t)= \rho_g(a(t),\;\dot{a}(t))\;.
\end{equation}
Therefore, the system of coupled equations (\ref{e18}), (\ref{e21}) and (\ref{e22}) allows us
to obtain $a(t)$ taking into account the back reaction induced by the
transformation of very long tensor perturbations into effective gravitational
waves.

Let us define
\begin{equation}
\label{e23}
A(t)\equiv \frac{a(t)}{a_2}\;,
\end{equation}
and
\begin{equation}
\label{e24}
a_2\equiv a(t_2)\;,
\end{equation}
\begin{equation}
\label{e25}
a_1\equiv a(t_1)\;,
\end{equation}
\begin{equation}
\label{e26}
H_1\equiv H(t_1)\;,
\end{equation}
\begin{equation}
\label{e27}
\sigma\equiv \frac{a_2}{a_1}\;,
\end{equation}

Then Eqs. (\ref{e8})--(\ref{e11}), (\ref{e18}), (\ref{e21}), and the expressions derived
for the Bogoliubov coeficients in \cite{gw1,gw3t}, lead to a lengthy set of equations that
enable us to evaluate $a(t)$ during the matter age. These equations take the form
\begin{equation}
\label{e28}
\dot{A}^2 =\frac{H_1^2}{\sigma^4}\;\frac{1}{A}\;\left [1+\kappa\;\left (\frac{H_1}{m_{Pl}}
\right )^2\;\frac{1}{A}\;G(t)\right ]\;,
\end{equation}
where $m_{Pl}$ is the Planck mass. The values of the constant $\kappa$ and of the
function $G(t)$ depends whether $\gamma =0$ (the initial era is a de Sitter one), or
$\gamma \neq 0$ (we will restrict our analysis to inflationary models, so that $\gamma < 2/3$).

For $\gamma = 0$:

\begin{equation}
\label{e29}
\kappa = \frac{2}{3\pi}\;,
\end{equation}
\begin{equation}
\label{e30}
G(t) = \left (1+\frac{1}{16\sigma^4}\right )\ln\tau+\frac{b_2^4}{2048\pi^4}(\tau^4-1)+
\frac{b_2^2}{32\pi^2}j(t)+\ln\left(\frac{b_2\sigma}{b_1}\right )+ c_0\;,
\end{equation}
\begin{equation}
\label{e31}
\tau\equiv \frac{H_1}{b_2\,\sigma^2\dot{A}}\;,
\end{equation}
\begin{equation}
\label{e32}
c_0\equiv 16\pi^4\left (\frac{1}{b_1^4}-\frac{1}{b_2^4\,\sigma^4}\right )\;,
\end{equation}
\begin{eqnarray}
\label{e33}
j(t)&\equiv& N_1(\sigma,\;\tau)\sin\theta_1+N_2(\sigma,\;\tau)\cos\theta_1+N_3(\sigma)
\cos\theta_2+N_4(\sigma)\sin\theta_2+\nonumber\\
& &\mbox{}+N_5(\sigma)J(t)\;,
\end{eqnarray}
\begin{equation}
\label{e34}
\theta_1(t)\equiv 4\pi \sigma (\sigma-1)\frac{\dot{A}}{H_1}\;,
\end{equation}
\begin{equation}
\label{e35}
\theta_2\equiv \frac{4\pi (\sigma-1)}{b_2\,\sigma}\;,
\end{equation}
\begin{equation}
\label{e36}
J(t)\equiv\int_{\theta_1(t)}^{\theta_2}\frac{\cos\theta}{\theta}d\theta\;,
\end{equation}
\begin{equation}
\label{e37}
N_1(\sigma,\tau)\equiv \frac{b_2}{4\pi}\left [\frac{(3\sigma+1)}{\sigma}\tau^3-
\frac{b_2(11\sigma^3+11\sigma^2-7\sigma+1)}{\pi}\tau\right ]
\end{equation}
\begin{equation}
\label{e38}
N_2(\sigma,\tau)\equiv \frac{1}{2}\left [\frac{(11\sigma^2+6\sigma-1)}{\sigma^2}
\tau^2-\frac{b_2^2}{8\pi^2}\tau^4
\right ]\;,
\end{equation}
\begin{equation}
\label{e39}
N_3(\sigma)\equiv \frac{1}{2}\left [ \frac{b_2^2}{8\pi^2}-\frac{(11\sigma^2+6\sigma-1)}
{\sigma^2}\right ]\;,
\end{equation}
\begin{equation}
\label{e40}
N_4(\sigma)\equiv \frac{b_2}{4\pi}\left [\frac{b_2(11\sigma^3+11\sigma^2-7\sigma+1)}{\pi}-
\frac{(3\sigma+1)}{\sigma}\right ]\;,
\end{equation}
\begin{equation}
\label{e41}
N_5(\sigma)\equiv -\frac{b_2(\sigma-1)}{\pi\sigma}(11\sigma^3+11\sigma^2-7\sigma+1)\;.
\end{equation}

On the above equations the free parameters $b_1$ and $b_2$ are related to the transitions
time scales $\Delta t_1$ and $\Delta t_2$ that are assumed to take place at $t_1$ and $t_2$,
respectively:
\begin{equation}
\label{e42}
\Delta t_1 = \frac{b_1}{H(t_1)}=\frac{b_1}{H_1}\;,
\end{equation}
\begin{equation}
\label{e43}
\Delta t_2 = \frac{b_2}{H(t_2)}=\frac{b_2}{H_2}\;.
\end{equation}
Thus, the larger $b_i$, the slower is the transition in comparison with the Hubble time at
$t_i$, $H_i^{-1}$. It is usual to take $b_1 \sim b_2 \sim 1$, but, as for the first
transition, there is no compelling reason to assume that it necessarily ocurred in a
time scale of the order of $H_1^{-1}$ in every model. For example, this transition may
occurr as fast as $10^{-4}H_1^{-1}$ in the `new' inflationary scenarios \cite{allen}.

When the first stage is not a de Sitter one, that is, for $\gamma<2/3$, $\gamma\neq 0$,
the constant $\kappa$ and the function $G(t)$ become
\begin{equation}
\label{e44}
\kappa =\frac{4}{3}\;,
\end{equation}
\begin{equation}
\label{e45}
G(t)=\frac{1}{\pi\sigma^3}G_1(t)+\frac{4}{|2m+1|^4}C_0\;,
\end{equation}
where
\begin{equation}
\label{e46}
m\equiv \frac{3}{2}\frac{(2-\gamma)}{(3\gamma-2)}<0\;,
\end{equation}
\begin{equation}
\label{e47}
C_0\equiv \frac{1}{4}\left (m-\frac{1}{2}\right )^2\left (m+\frac{1}{2}\right )^2
\int_{y_2}^{y_1}B_m(y)dy+D(y_1)-D(y_2)\;,
\end{equation}
\begin{equation}
\label{e48}
y_1\equiv \frac{\pi|2m+1|}{b_1}\;,
\end{equation}
\begin{equation}
\label{e49}
y_2\equiv \frac{\pi|2m+1|}{b_2\,\sigma}\;,
\end{equation}
\begin{equation}
\label{e50}
D(y)\equiv f_1(m,y)B_m(y)+f_2(m,y)B_{m+1}(y)+f_3(m,y)C_m(y)+\frac{y^4}{\pi}\;,
\end{equation}
\begin{eqnarray}
\label{e51}
f_1(m,y)& \equiv & \frac{1}{4}\left [y^5+\frac{1}{2}\left (m+\frac{1}{2}\right )\left
(m+\frac{3}{2}\right )y^3-\right.\nonumber\\
& &\mbox{}\left.-\frac{1}{2}\left (m-\frac{1}{2}\right )^2\left (m+\frac{1}{2}\right )
y\right ]\;,
\end{eqnarray}
\begin{equation}
\label{e52}
f_2(m,y)\equiv \frac{1}{4}\left [y^5+\frac{1}{2}\left (m+\frac{1}{2}\right )\left (
m-\frac{1}{2}\right )y^3\right ]\;,
\end{equation}
\begin{equation}
\label{e53}
f_3(m,y)\equiv -\frac{1}{2}\left (m+\frac{1}{2}\right )\left [y^4+\frac{1}{2}\left
(m-\frac{1}{2}\right )^2y^2\right ]\;,
\end{equation}
\begin{equation}
\label{e54}
B_m(y)\equiv J_m^2(y)+Y_m^2(y)\;,
\end{equation}
\begin{equation}
\label{e55}
C_m(y)\equiv J_m(y)J_{m+1}(y)+Y_m(y)Y_{m+1}(y)\;.
\end{equation}
$J_m$ and $Y_m$ are, respectively, the Bessel functions of the first and second kinds
of order $m$, and
\begin{eqnarray}
\label{e56}
G_1(t)& \equiv & \frac{1}{32\sigma}\ln\tau+P(m,\sigma,b_2)\left [\frac{\pi^2}{b_2^3(-2m-3)}
\left (\tau^{-2m-3}-1\right )+\right.\nonumber\\
& &\mbox{}
\left.+\frac{b_2}{512\pi^2(1-2m)}\left (\tau^{1-2m}-1\right )+\frac{1}{128\pi}f(t)\right ]\;,
\end{eqnarray}
\begin{equation}
\label{e57}
P(m,\sigma,b_2)\equiv \left (\frac{2\,b_2\,\sigma}{\pi |2m+1|}\right )^{-2m}\,\Gamma^2(-m)
|2m+1|\;,
\end{equation}
\begin{eqnarray}
\label{e58}
f(t)&\equiv& P_1(m,\sigma)\tau^{-2m}\sin\theta_1 +\left [\frac{4\pi}{b_2}P_2(m,\sigma)
\tau^{-2m-1}\,-\right.\nonumber\\
& &\mbox{}\left.-\frac{b_2}{4\pi (1-2m)}\tau^{1-2m}\right ]\cos\theta_1 +
P_3(m,\sigma,b_2)\cos\theta_2-\nonumber\\
& &\mbox{}- P_1(m,\sigma)\sin\theta_2+P_4(m,\sigma)I(t)\;,
\end{eqnarray}
\begin{equation}
\label{e59}
I(t)\equiv \int_{\theta_1(t)}^{\theta_2}\frac{\sin\theta}{\theta^{-2m-1}}d\theta\;,
\end{equation}
\begin{equation}
\label{e60}
P_1(m,\sigma)\equiv \frac{3\sigma-2m(1+2\sigma)}{(-2m)(1-2m)\sigma}\;,
\end{equation}
\begin{equation}
\label{e61}
P_2(m,\sigma)\equiv \frac{(\sigma-1)}{\sigma^2|2m+1|}\left [\frac{3\sigma-2m(1+2\sigma)}
{(-2m)(1-2m)}+\frac{2\sigma(\sigma+1)}{(\sigma-1)}\right ]\;,
\end{equation}
\begin{equation}
\label{e62}
P_3(m,\sigma,b_2)\equiv \frac{b_2}{4\pi (1-2m)}-\frac{4\pi}{b_2}P_2(m,\sigma)\;,
\end{equation}
\begin{eqnarray}
\label{e63}
P_4(m,\sigma)&\equiv & -\frac{1}{\sigma |2m+1|}\left [\frac{4(\sigma-1)}{|2m+1|}\right ]
^{-2m}\left [\frac{3\sigma-2m(1+2\sigma)}{(-2m)(1-2m)}+\frac{2\sigma(\sigma+1)}{(\sigma-1)}
+\right.\nonumber\\
 & &\mbox{}\left.+\frac{2\sigma^2|2m+1|}{(\sigma-1)^2}\right ]\;,
\end{eqnarray}
and $\Gamma$ is the gamma function.
\section{NUMERICAL RESULTS AND CONCLUSIONS}
\label{s4}
We are interested in evaluating $a(t)$ after $t_2$, hence the initial conditions are
\begin{equation}
\label{e64}
A(t_2)=1
\end{equation}
and
\begin{equation}
\label{e65}
\dot{A}(t_2)=\frac{H_1}{\sigma^2}\;.
\end{equation}
Note that, besides $b_1$ and $b_2$, there are three free parameters:
\begin{equation}
\label{e66}
h_1 \equiv H_1/m_{Pl}\;,
\end{equation}
$\sigma$, whcih we relate to the duration of the radiation era $t_2-t_1$ through
\begin{equation}
\label{e67}
\sigma^2= 1+ 2H_1(t_2-t_1)\;,
\end{equation}
and
$m$ (or, equivalently, $\gamma$).

We have integrated Eq. (\ref{e28}) numerically for different
values of the parameters $b_1$, $b_2$, $h_1$, $\gamma$, and
$t_2-t_1$. We have taken as characteristic values $b_1=1$,
$b_2=1$, $h_1= 10^{-50}$ and $0\leq\gamma\leq 0.6\,$. In Figure 1
we show the solutions for five values of $t_2-t_1$, namely
$3\times 10^6\,{\rm yr}$ (curve a), $10^6\,{\rm yr}$ (curve b),
$3\times 10^5\,{\rm yr}$ (curve c), $10^5\,{\rm yr}$ (curve d), and
$3\times 10^4\,{\rm yr}$ (curve e). The time is measured in units
of the time of the beginning of the matter era $t_2$. The
important thing to notice is that the solutions are insensitive to
all parameters, except $t_2-t_1$. We see that different values of $t_2-t_1$
change the behaviour of $A(t)$ near $t_2$. All the curves go
assymptotically as $t^{2/3}$, which is represented by the dotted
curve. However, the net effect of decreasing the duration of the
radiation era is to change the present value of $a(t)/a_2$.

We are led to conclude that the transformation of very long tensor
perturbations into effective gravitational waves does not change
the expansion dynamics during the matter epoch. This is in
contrast to what happens during the radiation era, as it has been
shown in Ref. \cite{gw6}. Therefore, the Sahni's conjecture
\cite{sahni} will hold only under very restrictive conditions.

We must remark, however, that, in the model analysed above,
we have assumed a standard evolution
from a radiation phase with $a(t)\propto t^{1/2}$ to a dust era
with $a(t)\propto t^{2/3}$. We have not taken into account the
modifications in the dynamics generated by the gravitons' back
reaction during the radiation dominated period \cite{gw6}. A more realistic
model would consider these modifications and {\em then} a
transition to a matter dominated era. The numerical work becomes
much more involved in this case. Moreover, primordial nucleosynthesis and the
data related to the anisotropy of the cosmic microwave background will
place severe constraints on the relevant parameters.
These questions are presently under investigation.
\acknowledgments
The authors would like to thank  the Brazil Research Council (CNPq) for
financial support.

\begin{figure}
\vspace{.2in}
\centerline{\psfig{figure=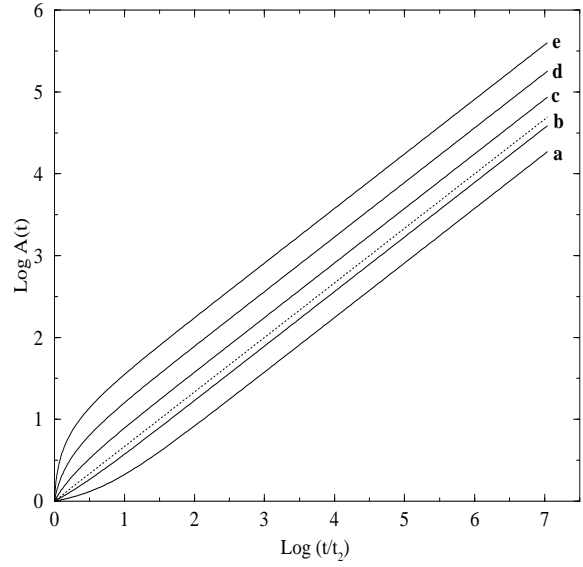,width=3truein,height=3truein}
\hskip 0.1in}
\caption{A logarithmic plot of the quantity $A=a/a_2$ as a function of time
[solution of Eq. (28)]
for $b_1=1$, $b_2=1$, $h_1= 10^{-50}$, $\gamma = 0.5$, and five
values of $t_2-t_1$:
$3\times 10^6\,{\rm yr}$ (curve a), $10^6\,{\rm yr}$ (curve b),
$3\times 10^5\,{\rm yr}$ (curve c), $10^5\,{\rm yr}$ (curve d), and
$3\times 10^4\,{\rm yr}$ (curve e). The dotted curve corresponds
to $a(t) \propto  t^{2/3}$ and the log is to base 10. The solutions
are insensitive to all parameters, except $t_2-t_1$.}
\end{figure}

\begin{references}
\bibitem{gw6} M. R. de Garcia Maia, J. C. Carvalho, and J. S. Alcaniz, Phys. Rev. D
{\bf 56}, 6351 (1997).
%
\bibitem{allen} B. Allen, Phys. Rev. D {\bf 37}, 2078 (1988).
%
\bibitem{gw1} M. R. G. Maia, Phys. Rev. D {\bf 48}, 647 (1993).
%
\bibitem{gw3t} M. R. G. Maia and J. D. Barrow, Phys. Rev. D {\bf 50}, 6262 (1994);
M. R. G. Maia, Ph. D. thesis, University of Sussex, 1994.
%
\bibitem{gw4} M. R. G. Maia and J. A. S. Lima, Phys. Rev. D {\bf 54}, 6111 (1996).
%
\bibitem{abbott} L. F. Abbott and D. D. Harari, Nucl. Phys. {\bf B264}, 487 (1986).
%
\bibitem{birrel} N. D. Birrel and P. C. Davies, {\em Quantum Fields in Curved Space}
(Cambridge University Press, Cambridge, England, 1982).
%
\bibitem{fulling} S. A. Fulling, {\em Aspects of Quantum Field Theory in Curved Specetime}
(Cambridge University Press, Cambridge, England, 1989).
%
\bibitem{zeldovich} Ya. B. Zel'dovich and I. D. Novikov, Astron. Zh. {\bf 46}, 960 (1970)
[Sov. Astron. {\bf 13}, 754 (1970)]; {\em The Structure and Evolution of the Universe}
(The University of Chicago Press, Chicago, 1983), Vol. 2.
%
\bibitem{prigogine} I. Prigogine, J. Geheniau, E. Gunzig, and P. Nardone, Gen. Relativ.
Gravit. {\bf 21}, 767 (1989).
%
\bibitem{calvao} M. O. Calv\~ao, J. A. S. Lima, and I. Waga, Phys. Lett, A {\bf 162}, 223
(1992).
%
\bibitem{sahni} V. Sahni, Phys. Rev. D {\bf 42}, 453 (1990).
%
\bibitem{vilenkin84} A. Vilenkin, Phys. Rev. Lett. {\bf 53}, 1016
(1984).
%
\bibitem{gasperini92} M. Gasperini and G. Veneziano, Phys. Lett. B
{\bf 277}, 256 (1992).
%
\bibitem{kolb89} E. W. Kolb, Astrophys. J. {\bf 344}, 543 (1989).
%
\bibitem{akdeniz94} K. G. Akdeniz, M. Arik, and E. Rizaoglu, Phys.
Lett. B {\bf 321}, 329 (1994).
%
\bibitem{wesson92} P. S. Wesson, Phys. Ess. {\bf 5}, 561 (1992).
%
\bibitem{gri93} L. P. Grishchuk, Phys. Rev. D {\bf 48}, 5581 (1993).
%
\bibitem{gw5} D. M. Tavares and M. R. G. Maia, Phys. Rev. D {\bf 57}, 2305 (1998).
%
\bibitem{kT} E. Kolb and M. Turner, {\em The Early Universe} (Addison-Wesley,
Redwood City, CA, 1990).
%
\bibitem{carr} B. J. Carr, Astron. Astrophys. {\bf 89}, 6 (1980).
%
\bibitem{gri}
L. P. Grishchuk, Zh. \'Eksp. Teor. Fiz. {\bf 67}, 825 (1975) [Sov. Phys. JETP
{\bf 40}, 409 (1975)];
Lett. Nuovo Cimento {\bf 12},60 (1975);
Ann. (N. Y.) Acad. Sci. {\bf 302}, 439 (1977);
Usp. Fiz. Nauk. {\bf 121}, 629 (1977) [Sov. Phys. Usp. {\bf 20}, 319 (1977)];
L. P. Grishchuk and Y. V. Sidorov, Phys. Rev. D {\bf 42}, 3413 (1990);
L. P. Grishchuk, Class. Quantum Grav. {\bf 10}, 2449 (1993).
%
\end{references}
\end{document}